\documentclass[aps,prd,twocolumn,notitlepage,showpacs,showkeys,nosuperscriptaddress,nofootinbib,10pt]{revtex4-2}

\def \be {\begin{equation}} 
\def \ee {\end{equation}} 
\def \bea {\begin{eqnarray}} 
\def \eea {\end{eqnarray}} 

\usepackage{physics}
\usepackage{commath}
\usepackage{url}
\usepackage{bm}
\usepackage{braket}
\usepackage{accents}
\usepackage{graphicx}
\graphicspath{{figs/}}

\usepackage{dcolumn}
\usepackage{bm}
\usepackage{epsfig} 
\usepackage{amsfonts}
\usepackage{amsmath}
\usepackage{amssymb}
\usepackage[usenames]{color}
\usepackage[dvipsnames]{xcolor}
\usepackage[unicode, colorlinks=true, linkcolor=linkcolor, citecolor=linkcolor, filecolor=linkcolor,urlcolor=linkcolor, pdfusetitle]{hyperref}
\usepackage{listings}
\usepackage[T1]{fontenc}
\usepackage{caption}

\hypersetup{colorlinks,citecolor=blue,linkcolor=blue,urlcolor=blue}
\hypersetup{final=true}

\begin{document}

\title{Model-independent cosmological inference post DESI DR1 BAO measurements}

\author{Purba Mukherjee}
\email{pdf.pmukherjee@jmi.ac.in}

\author{Anjan Ananda Sen}
\email{aasen@jmi.ac.in}

\affiliation{Centre for Theoretical Physics, Jamia Millia Islamia, New Delhi-110025, India}

\bigskip

\begin{abstract}
\vskip 0.1cm

In this work, we implement Gaussian process regression to reconstruct the expansion history of the universe in a model-agnostic manner, using the Pantheon-Plus SN-Ia compilation in combination with two different BAO  measurements (SDSS-IV and DESI DR1). In both the reconstructions, the $\Lambda$CDM model is always included in the 95\% confidence intervals. We find evidence that the DESI LRG data at $z_{\text{eff}} = 0.51$ is not an outlier within our model-independent framework. We study the $\mathcal{O}m$-diagnostics and the evolution of the total equation of state (EoS) of our universe, which hint towards the possibility of a quintessence-like dark energy scenario with a very slowly varying EoS, and a phantom-crossing in higher $z$. The entire exercise is later complemented by considering two more SN-Ia compilations - DES-5YR and Union3 - in combination with DESI BAO. Reconstruction with the DESI BAO + DES-5YR SN data sets predicts that the $\Lambda$CDM model lies outside the 3$\sigma$ confidence levels, whereas with DESI BAO + Union3 data, the $\Lambda$CDM model is always included within 1$\sigma$. We also report constraints on $H_0 r_d$ from our model-agnostic analysis, independent of the pre-recombination physics. Our results point towards an $\approx$ 2$\sigma$ discrepancy between the DESI + Pantheon-Plus and DESI + DES-5YR data sets, which calls for further investigation.

\end{abstract}

\vskip 1.0cm



\maketitle

%\vskip 1.0cm

%%%%%%%%%%%%%%%%%%%%%%%%%%%%%%%%%%%%%%%%%%%%%%%%%%%

\section{Introduction \label{sec:intro}}

Recent advancements in the precision of cosmological observations have significantly bolstered support for the $\Lambda$CDM model, which has been highly effective in explaining a wide range of phenomena \cite{SupernovaSearchTeam:1998fmf, SupernovaCosmologyProject:1998vns, Blanchard:2022xkk, Peebles:2024txt}. These include fluctuations in the temperature and polarization of the cosmic microwave background (CMB) \cite{Planck:2015bue, Planck:2018vyg, ACT:2020gnv, Tristram:2023haj}, the large-scale structure of the universe \cite{BOSS:2014hhw, BOSS:2016wmc, eBOSS:2020yzd, Addison:2017fdm}, and the distance-redshift relation of type Ia supernovae (SNIa) \cite{SDSS:2014iwm, Pan-STARRS1:2017jku, Brout:2022vxf}. Despite its success, this 6-parameter description faces several theoretical challenges (such as the elusive nature of dark matter \cite{dark_matter, LUX:2016ggv}, the cosmological constant and the cosmic coincidence problem \cite{Weinberg:1988cp, Sahni:1999gb, Carroll:2000fy, Padmanabhan:2002ji, Peebles:2002gy}) as well as observational inconsistencies \cite{Hazra:2013dsx, Bull:2015stt, Verde:2019ivm, Riess:2019qba, Perivolaropoulos:2021jda,  Efstathiou:2024dvn}. Chief among these concerns is the current >5$\sigma$ discrepancy between the Cepheid-calibrated local distance ladder measurements \cite{Riess:2021jrx} of the Hubble constant $H_0$ when compared to CMB-inferred values \cite{Planck:2018vyg}, alongside $\sim 2-2.5\sigma$ tension in the amplitude of matter fluctuations $S_8$ derived from early CMB \cite{Planck:2018vyg} vs weak lensing surveys \cite{Heymans:2020gsg, DES:2021wwk, Li:2023tui}. Additionally, deep-space observations from the James Webb Space Telescope have revealed the presence of massive galaxies at very high redshifts ($z \sim 15$) \cite{Labbe:2022ahb, Boylan-Kolchin:2022kae}. While investigations into potential systematic errors contributing to these tensions will continue to thrive, it has motivated the need to explore new physics by revisiting the foundations of the concordance model or by introducing exotic components into the cosmological framework \cite{DiValentino:2020zio, DiValentino:2020vvd, DiValentino:2021izs, Shah:2021onj, Schoneberg:2021qvd, Kamionkowski:2022pkx, Abdalla:2022yfr, Vagnozzi:2023nrq, Hu:2023jqc}. 

The recent measurement of baryon acoustic oscillations (BAO) from the Dark Energy Spectroscopic Instrument (DESI)  has garnered significant attention, particularly its implications for dark energy behaviour. When combined with Planck CMB and SNIa observations, the DESI BAO measurements, indicate a greater degree of tension with the concordance model \cite{desi_paper6}. DESI suggests a preference for phantom behaviour of dark energy ($w<-1$) across a significant redshift range, implying a potential crossing of the phantom divide at lower redshifts, based on a spatially-flat $w_0 w_a$CDM cosmological parameterization \cite{Chevallier:2000qy, Linder:2002et} for dark energy, modelled as a fluid with a dynamical equation of state parameter $w(z)$. Although these results are intriguing, the CPL parametrization serves as a phenomenological ansatz rather than a physically consistent model for dark energy. In light of these findings, the scientific community has begun to explore various alternative scenarios (see Ref. \cite{eoin_desi, liddle_desi, Park:2024jns, Wolf:2023uno, Wolf:2024eph, Wang:2024hks, wang_desi, Calderon:2024uwn, DESI:2024kob, Gialamas:2024lyw, Pang:2024qyh, Jiang:2024viw, Jiang:2024xnu, Dinda:2024kjf, Dinda:2024ktd, Bhattacharya:2024hep, RoyChoudhury:2024wri}). 

The goal of this paper is to investigate the implications of these developments through a cosmological data-driven reconstruction of the Universe's evolutionary history, employing Gaussian process (GP) regression \cite{10.7551/mitpress/3206.001.0001, Holsclaw:2011wi, Seikel:2012uu, Shafieloo_2012}. In the absence of a viable cosmological model, we attempt to reconstruct the Hubble parameter $H$ and total energy density $\rho_t$ from current SN-Ia and BAO observations in a non-parametric and model-agnostic manner, thereby aiming to enhance our understanding of the nature of dark energy and assessing the potential need for new physics in light of current observational evidence.

This paper is organized as follows: Sec. \ref{sec:theo} outlines the details of the underlying theoretical framework. In Sec. \ref{sec:methodology}, we describe the methodology of our analysis for reconstruction, followed by a discussion of the results in Sec. \ref{sec:results}. Finally, in Sec. \ref{sec:conclusion} we summarize our findings and make some concluding remarks. We have also included a comprehensive overview of the reconstruction technique, Gaussian processes regression, and a description of the observational data sets, SNIa and BAO, in appendices \ref{sec:gp} and \ref{sec:data} respectively.

\section{Theoretical Framework \label{sec:theo}}

We start with spatially flat FRW metric in comoving coordinates,

\begin{equation} \label{eq:frw_metric}
\mathrm{d} s^2 = \mathrm{d} t^2 {-} a^{2}(t) \left[\mathrm{d}r^2 + r^2 \mathrm{d}\theta^2 + r^2 \sin^{2}\theta \mathrm{d}\phi^2\right], 
\end{equation}

\noindent
where $a(t)$ is the scale factor for the expanding Universe. In an ever-expanding Universe, $a(t)$ is a monotonic function of time and can be related to the cosmological redshift as $1+z = 1/a(t)$ where we have normalised the present value of the scale factor $a_{0} =1$ (in rest of the paper, quantities with subscript `$0$' will denote values at present). Given that in an expanding Universe, the redshift $z$ can also play the role of time, one can write the FRW metric in the following way (first considered by Choudhury \& Padmanabhan\cite{tirth_paddy}):
\begin{equation} \label{eq:mod_metric}
\mathrm{d}s^2 = \frac{1}{(1+z)^2}\left[\frac{\mathrm{d}z^2}{H^2 (z)} - \left(\mathrm{d}r^2+ r^2 \mathrm{d}\theta^2 + r^2 \sin^{2}\theta \mathrm{d}\phi^2 \right) \right],
\end{equation}

\noindent
where $H(z) = \frac{\dot{a}}{a}$ is the Hubble parameter and `overdot' represents derivative with respect to the time coordinate. Equation \eqref{eq:mod_metric} shows that in a spatially flat expanding FRW Universe, the only unknown function is the Hubble parameter $H(z)$. 

Any cosmological observation related to the geometry of the Universe (e.g SNIa measuring the luminosity distance \cite{SupernovaSearchTeam:1998fmf, SupernovaCosmologyProject:1998vns} or BAO measuring the angular diameter distance \cite{Bassett:2009mm}, etc.) can only constrain the parameter $H(z)$. Moreover, as $H(z)$ is related to the total energy density of the Universe $\rho_{t} (z)$ through the Einstein equation, such geometric observations tell us about $\rho_{t} (z)$ only and are agnostic to how $\rho_{t} (z)$ is distributed among different kind of components in the Universe (except for models in which the different components are interacting \cite{DiValentino:2017iww, Pan:2023mie, Giare:2024smz, Shah:2024rme}). Specifically, a certain constrained form for $H(z)$ can be represented by different dark energy or modified gravity models but cosmological observations that probe the FRW metric can not distinguish among all such models. Moreover, observations like CMB primarily probe the very early Universe when the dark energy effect is negligible \cite{Novosyadlyj:2013nya, Hazra:2013dsx, Bernal:2016gxb} (except the recently speculated early dark energy models to solve the Hubble Tension \cite{Poulin:2018cxd, Kamionkowski:2022pkx, Poulin:2023lkg}). Its dependence on the late Universe comes through the acoustic scale, which again can probe only $H(z)$ and is insensitive to various decomposition of $\rho_{t} (z)$. 

In this regard, if we consider the current Hubble Tension between SH0ES \cite{Riess:2021jrx} and Planck 2018 \cite{Planck:2018vyg} measurements for $H_{0}$ (assuming that the cause of this tension is not due to unknown systematics in either SH0ES or Planck 2018 observations), we can say that SH0ES predicts $\approx$ $9\%$ higher value for $H_0$, which corresponds to $\approx$ 19\% more total energy density $\rho_{t} (z)$ in the present Universe, compared to the Planck 2018 prediction. This enhancement in $\rho_{t} (z)$ at present may be due to (i) the cumulative increase due to new physics during the entire period of Universe evolution (from matter-radiation equality to the present day), (ii) enhancement due to new physics in the early Universe, or (iii) new physics in the late Universe (see \cite{Vagnozzi:2023nrq, Hu:2023jqc, DiValentino:2021izs, Shah:2021onj, Abdalla:2022yfr, Akarsu:2024qiq, Wang:2024hks}). More specifically, if the enhancement in $\rho_{t} (z)$ today is because of enhancement in $\rho_{t} (z)$ initially due to new physics in the early universe, that effect will be felt not only at present but also at every $z$, in particular at redshifts probed by SNIa and BAO observations which gives very tight constraints on $H(z)$ and hence on $\rho_{t}(z)$ in the redshift range $0.1\leq z \leq 2.5$. Hence, it is natural to ask whether we see such enhancement in $\rho_{t}(z) $ (compared to $\Lambda$CDM model as constrained by Planck-2018) in such redshift range from SNIa and BAO observations. In this paper, we study this through a cosmological model-independent reconstruction of $H(z)$ (and hence $\rho_{t} (z)$) using GP regression \cite{10.7551/mitpress/3206.001.0001, Seikel:2012uu, Holsclaw:2011wi, Shafieloo_2012, Mukherjee:2022lkt, Banerjee:2023evd, Banerjee:2023rvg, Favale:2023lnp} utilizing the latest SNIa and BAO datasets.

\begin{table}[t!]
{\renewcommand{\arraystretch}{1.15} \setlength{\tabcolsep}{7.5 pt} \centering
\begin{tabular}{l | c  c  c  c }
%        \hline 
        \hline 
        Kernels &  RBF & RQD & M72 & M92 \\
        \hline 
        SDSS+Pantheon-Plus &  0.878  &	 0.878	&   0.878	&   0.878   \\
        DESI+Pantheon-Plus &  0.88	 &   0.882	&   0.879	&   0.88    \\
        DESI+DES-5YR       &  0.889  &   0.89   &   0.889   &   0.889   \\
        DESI+Union3        &  0.962  &   1.091  &   0.933   &   0.947   \\
        \hline
%        \hline 
    \end{tabular}
    \caption{Reduced $\chi^2_{\text{min}}$ values obtained with different data sets.}
    \label{tab:kernel}
}
\end{table}

\begin{table}[t!]
{\renewcommand{\arraystretch}{1.15} \setlength{\tabcolsep}{17.5 pt} \centering
\begin{tabular}{l | c }
%        \hline 
        \hline 
        Parameters &  Priors \\
        \hline 
        $\log_{10} \sigma_f$ &  $\mathcal{U}[-5, 6]$    \\
        $\log_{10} l$ &  $\mathcal{U}[-5, 6]$	   \\
        $M_B$       &  $\mathcal{U}[-21, -18]$   \\
        $H_0 \, r_d$        &  $\mathcal{U}[8000, 12000]$  \\
        \hline
%        \hline 
    \end{tabular}
    \caption{Priors on the kernel hyperparameters and cosmological parameters for Bayesian analysis.}
    \label{tab:priors}
}
\end{table}

\begin{table*}[t!]
{\small \renewcommand{\arraystretch}{1.5} \setlength{\tabcolsep}{12 pt} \centering
	\begin{tabular}{l | c | c  | c | c}
			\hline 
			& $\log_{10}\sigma_f$ & $\log_{10}l$ & $M_B$ & $H_0  r_d$ [100 km/s] \\
			\hline 
			Pantheon-Plus & $1.019^{+0.454}_{-0.345}$ & $0.542^{+0.238}_{-0.210}$ &  $-19.35 \pm 0.016$ & - \\
			SDSS BAO &  $1.813^{+0.407}_{-0.298}$  & $0.405^{+0.203}_{-0.181}$ &  -  & 	$100.23 \pm 2.04$ \\
			DESI BAO & $1.949^{+0.423}_{-0.341}$ &  $0.501^{+0.202}_{-0.200}$ &  -  & 	$103.27 \pm 2.35$ \\
			SDSS BAO +  Pantheon-Plus & $4.174^{+0.426}_{-0.352}$  & $0.662^{+0.194}_{-0.183}$ &  $-19.35$  & $99.348^{+0.728}_{-0.714}$ \\
			DESI BAO + Pantheon-Plus & $4.190^{+0.389}_{-0.334}$ & $0.635^{+0.195}_{-0.187}$ &  $-19.35$  & $100.481^{+0.837}_{-0.825}$ \\
			\hline
	\end{tabular}
    \caption{Marginalized constraints on the kernel hyperparameters and cosmological parameters.}
    \label{tab:mcmc}
    }
\end{table*}

\begin{figure}
\centering
\begin{minipage}{0.235\textwidth}
\includegraphics[width=\textwidth]{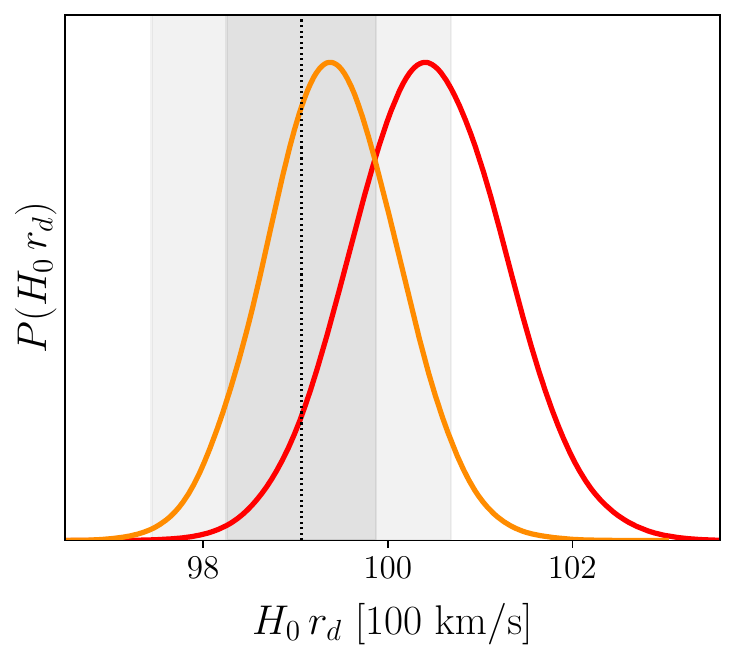}
\caption*{\small (a)} 
\end{minipage} 
\begin{minipage}{0.24\textwidth}
\includegraphics[width=\textwidth]{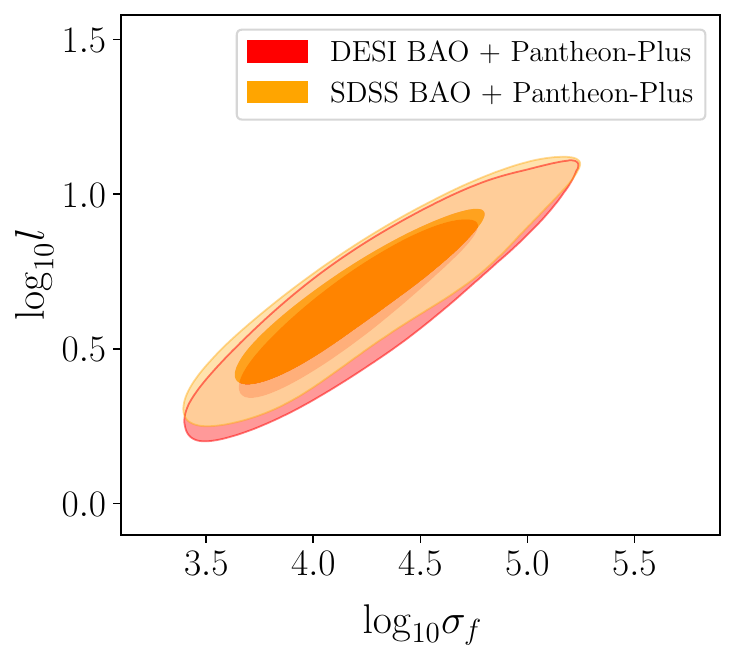}
\caption*{\small (b)} 
\end{minipage}
\begin{minipage}{0.235\textwidth}
\includegraphics[width=\textwidth]{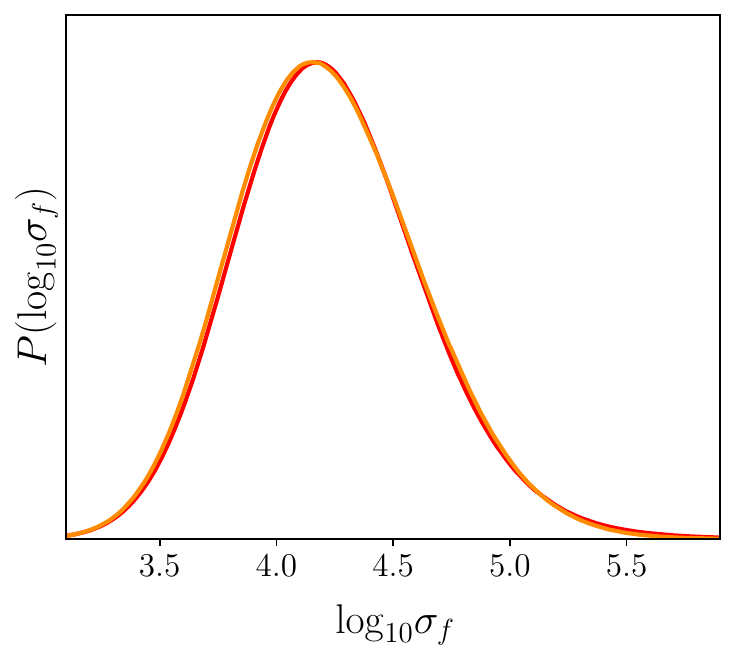}
\caption*{\small (c)} 
\end{minipage} 
\begin{minipage}{0.235\textwidth}
\includegraphics[width=\textwidth]{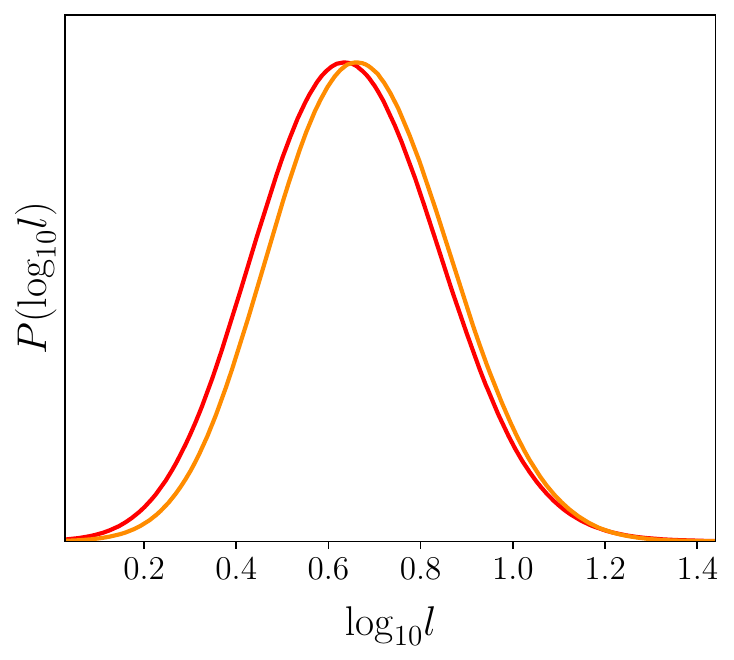}
\caption*{\small (d)} 
\end{minipage} 
%\hfill 	
\caption{(a) One-dimensional marginalized $H_0 r_d$ (in 100 km/s) posteriors for the SDSS+Pantheon-Plus vs DESI+Pantheon-Plus combinations. The black dashed lines and shaded regions represent the mean, 1$\sigma$ and 2$\sigma$ Planck 2018 \cite{Planck:2018vyg} constraints. (b) Two-dimensional contour plots for GP hyperparameter space assuming the Mat\'{e}rn 7/2 covariance function, along with one-dimensional marginalized posteriors for the hyperparameter (c) $\log_{10}{\sigma_f}$ and (d) $\log_{10}{l}$}.
\label{fig:hyperplot}
\end{figure}

\begin{figure*}[t!]
\centering
\begin{minipage}{0.485\textwidth}
\includegraphics[width=\textwidth]{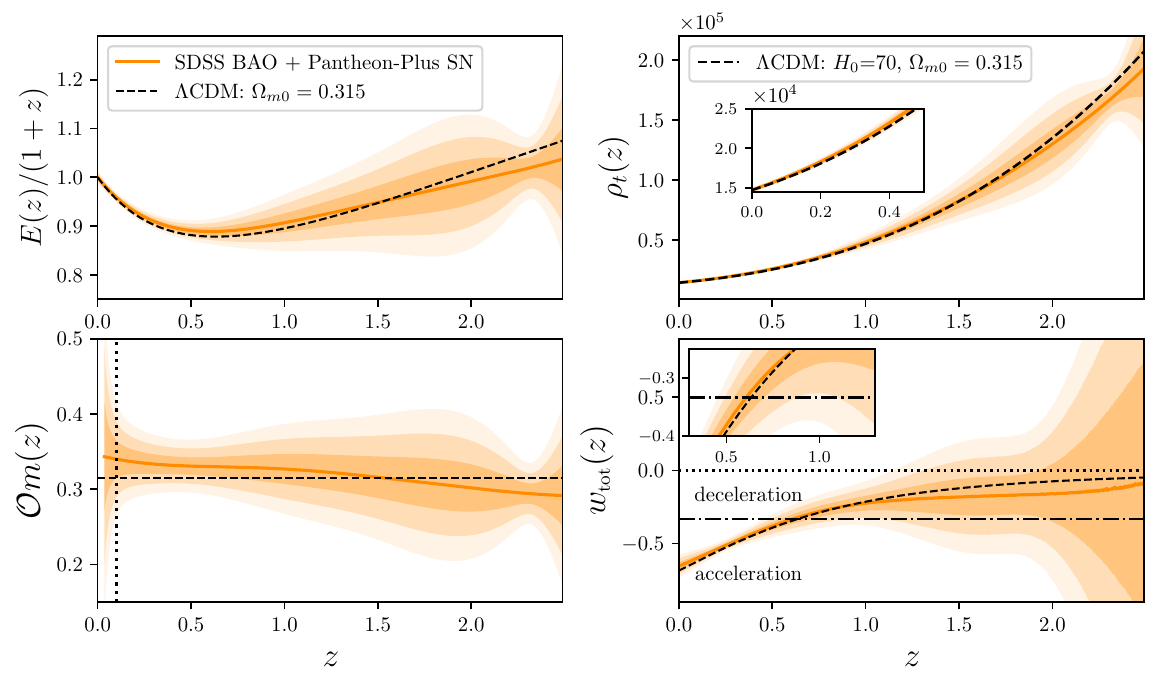}
\caption*{\small (a)} 
\end{minipage}
\begin{minipage}{0.485\textwidth}
\includegraphics[width=\textwidth]{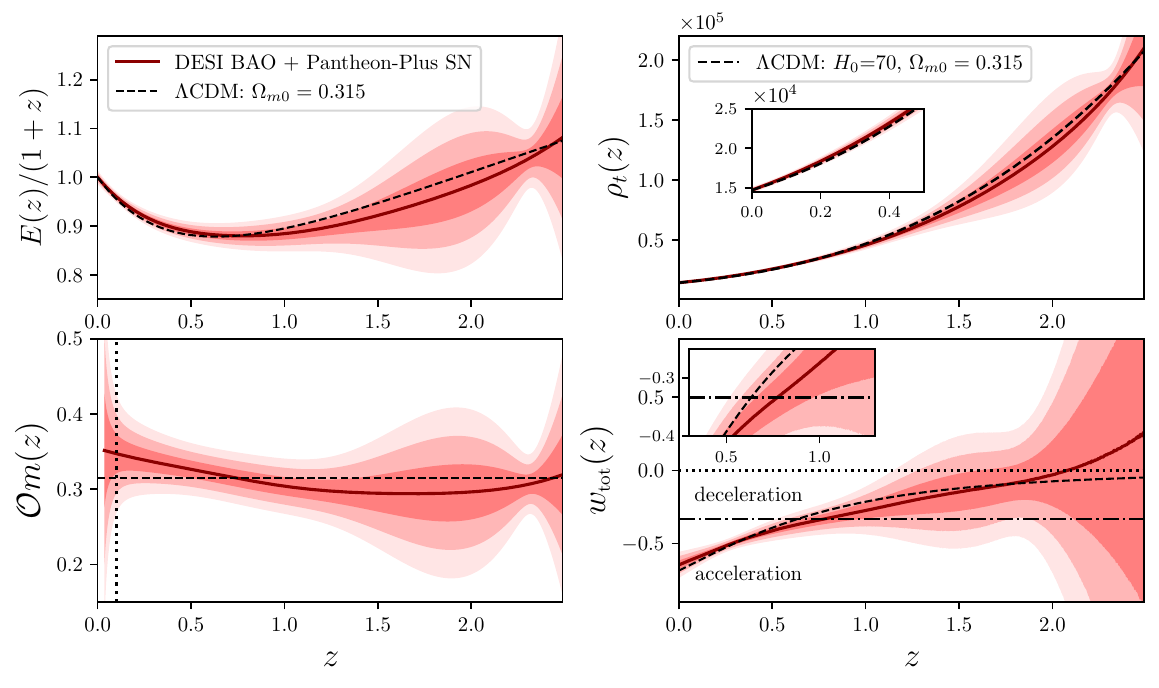}    
\caption*{\small (b)} 
\end{minipage} 
%\hfill 	
\caption{Plots for mean values of the reconstructed functions along with the 1$\sigma$, 2$\sigma$ and 3$\sigma$ uncertainties for the (a) SDSS BAO + Pantheon-Plus vs (b) DESI BAO +  Pantheon-Plus data sets. }
\label{fig:Ez_vs_z}
\end{figure*}

\section{Methodology \& Analysis \label{sec:methodology}}

To begin with, we consider two different BAO datasets, namely SDSS \cite{eBOSS:2020yzd} vs DESI DR1 \cite{desi_paper6}, in combination with the Pantheon-Plus \cite{Brout:2022vxf, Scolnic:2021amr} SN-Ia compilation, which is a high-precision and well-calibrated dataset covering a wide redshift range $0<z<2.26$, widely accepted in the cosmology community. We included both SDSS and DESI DR1 BAO measurements to make our findings more reliable. Even though DESI is newer and more comprehensive, using SDSS-IV BAO helps us cross-check our results, connect with the existing literature, and understand any differences between these two surveys. A GP regression is initially implemented on the Pantheon-Plus data employing the Mat\'{e}rn 7/2 kernel, where we have set the GP mean function to 0 for a cosmological model-independent analysis. The choice of this Mat\'{e}rn 7/2 kernel is made such that the reconstructed function is at least twice differentiable, without invoking additional smoothness such that the associated GP model corresponds to the maximum log-likelihood (Eq. \eqref{eq:loglike}). Table \ref{tab:kernel} shows that this particular kernel choice results in minimum values of the reduced $\chi^2$ i.e., $\tilde{\chi}^2_{\text{min}}$ \cite{Favale:2023lnp} (see Appendix \ref{sec:gp}). Besides the GP hyperparameters ($\sigma_f$ \& $l$), we keep the absolute magnitude $M_B$ of SN-Ia as an additional free parameter and transform the apparent magnitude $m$ to comoving distance $D_M$ measurements, such that 
\begin{equation}
D_M \left( z, \left\lbrace M_B \right\rbrace \right) = {10^{\frac{1}{5} \left[ m(z) - M_B - 25 \right]}}/(1+z) \, ,
\end{equation}
to reconstruct $D_M(z)$ and $D_M^{\prime}(z)$. The Hubble parameter can be computed as $H(z) = \frac{c}{D_M^\prime(z)}$, which is utilized to estimate $M_B$ such that $H(z=0)\equiv 70$ km Mpc$^{-1}$ s$^{-1}$. Our next step involves the inclusion of BAO data sets $D_M/r_d$, $D_H/r_d$ and the joint covariance matrix where $r_d$ is the sound horizon at drag-epoch. With the trained GP hyperparameters and best-fit value of $M_B \simeq -19.35$, we obtain predictions for $D_M(z_{_\text{BAO}})$ and $D_M^{\prime}(z_{_\text{BAO}})$ at the SDSS/DESI BAO redshifts $z_{_\text{BAO}}$ respectively. Instead of approximating the value of $r_d$ from standard early-time physics assumptions, we compute $H_0 r_d$ by maximizing the multivariate Gaussian likelihood for the BAO data sets and total covariance matrix obtained by combining the covariance matrix of the SDSS/DESI BAO and propagating the uncertainties of the reconstructed functions from the Pantheon-Plus data.  

Throughout this work, Bayesian MCMC analysis with \textsc{emcee} \cite{Foreman-Mackey:2012any} is adopted assuming uniform flat priors on the kernel hyperparameters, the SN-Ia absolute magnitude $M_B$ and $H_0 r_d$, summarized in Table \ref{tab:priors}. Table \ref{tab:mcmc} summarizes the marginalized constraints on $M_B$, $H_0 r_d$ and the kernel hyperparameters for the purpose of reconstruction. The best-fit constraints obtained on $H_0 r_d$ for both the combinations viz. $9934.88 ~ \pm ~72.73$ km/s with SDSS+Pantheon-Plus vs $10048.81 ~ \pm ~ 83.75$ km/s from DESI+Pantheon-Plus, shown in Fig. \ref{fig:hyperplot}(a), are completely consistent within the 1$\sigma$ confidence interval. The black dashed
lines together with the shaded regions represent the corresponding constraints from Planck 2018 \cite{Planck:2018vyg} mean, 1$\sigma$ and 2$\sigma$ confidence intervals. The allowed 2D-parameter space for the GP hyperparameter samples is shown in Fig. \ref{fig:hyperplot}(b) with \textsc{GetDist} \cite{Lewis:2019xzd}. The resulting marginalized posteriors for $\log_{10}\sigma_f$ and $\log_{10}l$ are shown in Fig. \ref{fig:hyperplot}(c) and Fig. \ref{fig:hyperplot}(d) respectively.

In what follows, we undertake a full model-agnostic reconstruction of the reduced Hubble parameter $E(z) = H(z)/H_0$ with the combined SDSS BAO + Pantheon-Plus and DESI BAO + Pantheon-Plus data. This ensures that there is no apparent tension between the BAO and SN-Ia data sets while attempting our joint analysis\footnote{See Ref. \cite{Calderon:2024uwn} for model-independent reconstruction of dark energy density with DESI DR1 data. To explore constraints on popular cosmological scenarios including inflation, modified gravity, annihilating dark matter and interacting dark energy, we direct readers to Ref. \cite{Wang:2024hks}.}.  Furthermore, we examine the $\mathcal{O}m$-diagnostics and the evolution of the total equation of state (EoS) of the universe. This investigation is enhanced by incorporating two additional SN-Ia compilations — DES-5YR \cite{DES:2024tys} and Union3 \cite{Rubin:2023ovl} — alongside DESI BAO data. We also investigate the impact of excluding the DESI DR1 LRG sample at an effective redshift of 0.51 from our analysis. Additionally, we provide constraints on $H_0 r_d$ from our model-agnostic analysis, independent of pre-recombination physics.

\section{Results \& Discussions \label{sec:results}}

We plot the reconstructed $E(z)$ scaled with $(1+z)$ in the top left panel of Fig. \ref{fig:Ez_vs_z}(a) \& Fig. \ref{fig:Ez_vs_z}(b), where the dashed line represents the $\Lambda$CDM model as constrained by Planck 2018. The solid curves are the mean of the reconstruction, whereas the shaded regions correspond to the 1$\sigma$, 2$\sigma$ and 3$\sigma$ confidence levels. We find that reconstructed $E(z)/(1+z)$ is consistent with Planck $\Lambda$CDM at 2$\sigma$. To visualize the presence of the late-time accelerating phase, it is more convenient to display the data as the phase portrait of the universe in the $\dot{a}- a$ plane as done by Choudhury \& Padmanabhan\cite{tirth_paddy}. Therefore, our focus lies on the evolution of the quantity $H_0^{-1} \, \dot{a} \equiv H_0^{-1} \, (a \, H)  = E \, (1+z)^{-1}$, as a function of the redshift. Because $E(z)/(1 + z)$ is directly proportional to $\dot{a}$, i.e. the expansion rate, its slope with redshift is directly proportional to minus the cosmic expansion acceleration ${\ddot{a}}$. The specific redshift at which the function $\frac{E(z)}{1+z}$ reaches a minimum indicates the transition redshift $z_t$ from a decelerated to an accelerated phase of the Universe's expansion. Therefore, for both cases studied, the range of variation in $E(z)$ shows that the current datasets predict a period of accelerated expansion at $z < 0.8$ within the 1$\sigma$ confidence level. 

We present the evolution of the total energy density of the universe $\rho_t(z)$, in the top right panels of Fig. \ref{fig:Ez_vs_z}(a) \& Fig. \ref{fig:Ez_vs_z}(b). Since our analysis assumes $H_0$ = 70 km Mpc$^{-1}$ s$^{-1}$, we plot the $\Lambda$CDM model predictions (with dashed lines) considering the same value of $H_0$ for comparison and it can be seen that such behaviour is within the $1\sigma$ region of our reconstructed behaviour. As $H^2(z)$ is proportional to $\rho_t(z)$, with Planck $\Lambda$CDM best-fit, the constant enhancement we require is about 4.29\% in $H(z=0)$ and 8.38\% in $\rho_t(z=0)$. Since this increase in $H_0$ can be obtained from early dark energy models, our results indicate that such an enhancement is consistent with our reconstructed behaviour for $\rho_t$. It further shows that any other late-time modification in the redshift range $0 < z < 2.5$ is allowed within our framework of reconstruction. Our findings confirm that the existing SN-Ia and BAO observations jointly cannot rule out either of these two possibilities. Therefore, given the current results, we can summarize that SNIa and BAO data sets cannot distinguish between any early or late-time new physics to resolve the $H_0$  tension, or search for unknown systematics in the SN-Ia data, such as revisiting the constancy of $M_B$, which has been assumed in our analysis and guides our determination of $H_0$ \cite{Dinda:2022jih, Gomez-Valent:2023uof, Mukherjee:2024akt, Camarena:2023rsd}.

\begin{figure}[t!]
\centering
\includegraphics[width=0.495\textwidth]{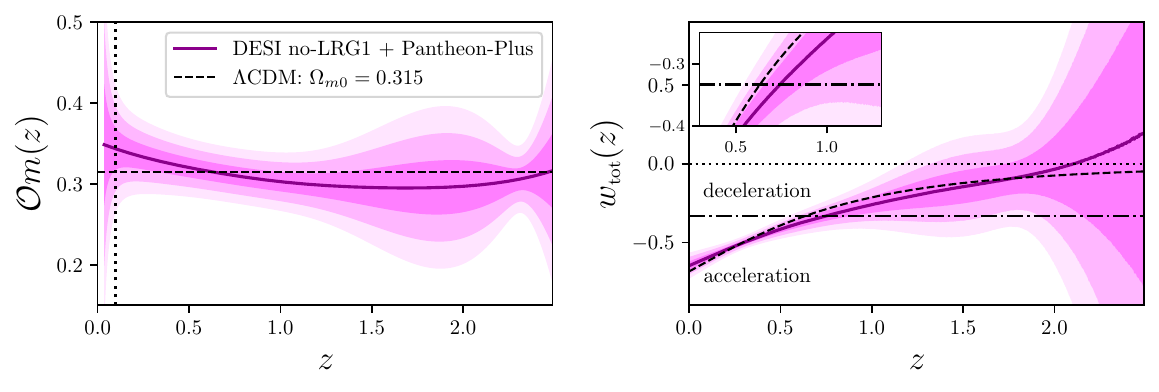} 
\caption{Plots for mean values of the reconstructed functions $\mathcal{O}m$ (left) and $w_{\text{tot}}$ (right) along with the 1$\sigma$, 2$\sigma$ and 3$\sigma$ uncertainties for the DESI BAO no-LRG1 + Pantheon-Plus SN data sets. }
\label{fig:compare_lrg}
\end{figure}

\begin{figure}[t!]
\centering
\begin{minipage}{0.485\textwidth}
\includegraphics[width=\textwidth]{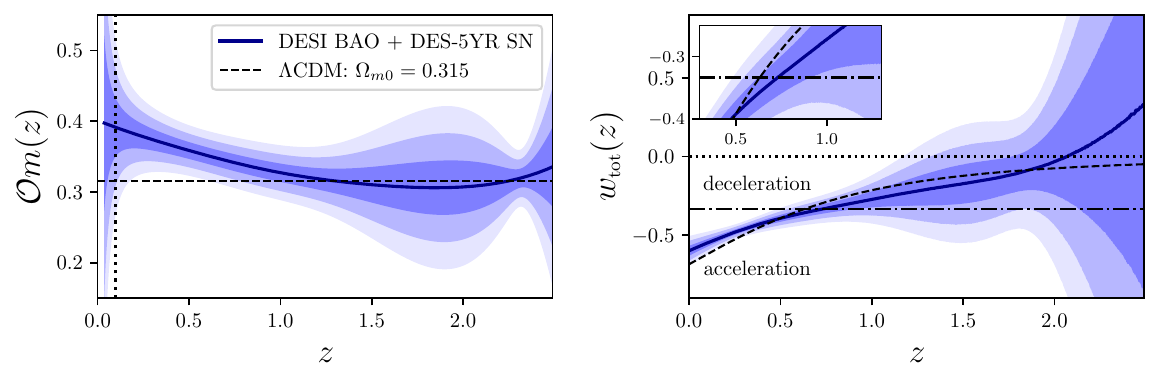}
\caption*{\small (a)} 
\end{minipage}
\begin{minipage}{0.485\textwidth}
\includegraphics[width=\textwidth]{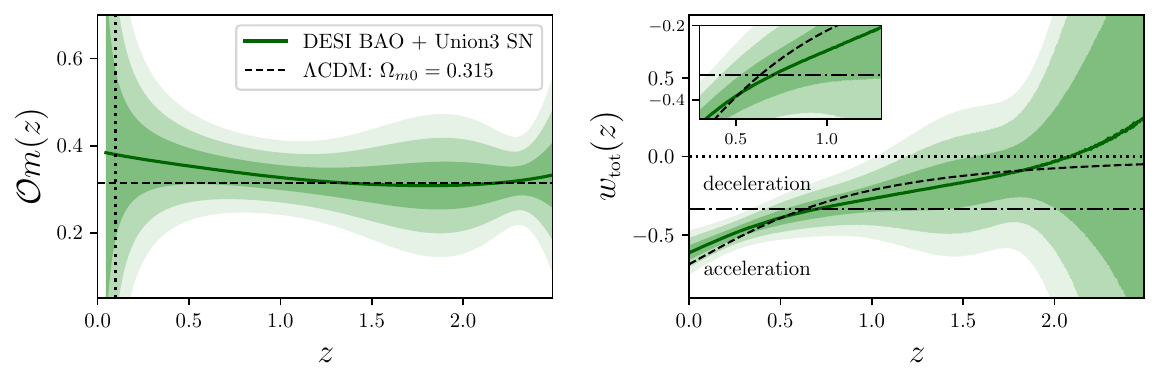}    
\caption*{\small (b)} 
\end{minipage} 
%\hfill 	
\caption{Plots for mean values of the reconstructed function $\mathcal{O}m$ (left) and $w_{\text{tot}}$ (right) along with the 1$\sigma$, 2$\sigma$ and 3$\sigma$ uncertainties for the (a) DESI BAO + DES-5YR vs (b) DESI BAO +  Union3 data sets. }
\label{fig:Ez_vs_z_2}
\end{figure}

We are interested in conducting a \textit{null test} to investigate whether dark energy behaves as a cosmological constant $\Lambda$, quintessence, phantom, or exhibits some form of exotic dynamical nature. For this purpose, we introduce the $\mathcal{O}m$ diagnostics \cite{Sahni:2008xx, Zunckel:2008ti, Shafieloo:2009hi}, defined as
\begin{equation}
\mathcal{O}m(z) = \frac{E^2(z)-1}{(1+z)^3 -1},
\end{equation} 
is an invaluable tool for probing the nature of dark energy without reliance on physical models. Different models have different evolutionary trajectories in the $z - \mathcal{O}m(z)$ plane.  As $\mathcal{O}m(z)$ relies exclusively on $E(z)$, inferred directly from observations, depending neither on $H'(z)$ nor on the cosmic growth factor, it offers an alternative route to differentiate between $\Lambda$CDM and dynamical DE models without directly invoking the DE EoS, in the absence of any compelling physical theory. For an FRW Universe, if the value of $\mathcal{O}m(z)$ remains constant across different $z$, then DE is exactly $\Lambda$ with $\mathcal{O}m(z) \equiv \Omega_{m0}$, the matter density parameter at the present epoch, and the underlying model is $\Lambda$CDM. Its utility lies in detecting deviations from this constancy: the slope of $\mathcal{O}m(z)$ can distinguish between different dark energy behaviours regardless of the value of $\Omega_{m0}$, even if $\Omega_{m0}$ is not accurately known. The degeneracy between evolving DE and $\Lambda$CDM can be broken by studying the behaviour of $\mathcal{O}m(z)$ at different redshifts. So, any possible deviation of $\mathcal{O}m(z)$ from $\Omega_{m0}$ can be used to draw insights into the dynamics of the Universe. For instance, in the case of a non-evolving EoS for dark energy, a positive slope of $\mathcal{O}m(z)$ suggests phantom ($w < -1$) while a negative slope indicates quintessence ($w > -1$) like behaviour \cite{Sahni:2008xx}. 

We plot the $\mathcal{O}m$ diagnostics as a function of redshift $z$ in the bottom left panels of Fig. \ref{fig:Ez_vs_z}(a) \& Fig. \ref{fig:Ez_vs_z}(b). The plot for DESI+Pantheon-Plus reveals a non-monotonic evolution of $\mathcal{O}m$, where the mean reconstructed curve exhibits a negative slope for $z<1$, remains nearly constant at $1.5<z<2$, and eventually becomes positive as $z > 2.3$. With SDSS+Pantheon-Plus, the slope of the reconstructed curve remains negative throughout the redshift range $0<z<2.5$. These subtle patterns, particularly the characteristic negative slope in the evolutionary trajectories of $\mathcal{O}m$, obtained in a model-agnostic way, indirectly suggest that dark energy exhibits quintessence-like behaviour at late times. Nonetheless, the $\Lambda$CDM model assuming Planck best-fit is included at the 2$\sigma$ confidence level. Figs. \ref{fig:Ez_vs_z}(a) \& \ref{fig:Ez_vs_z}(b) show that the reconstructed $\mathcal{O}m$ values are not well constrained at lower redshifts $z < 0.1$ (see dotted lines). The predictive power of \(\mathcal{O}m\) diagnostics at low $z$ is affected by the limited availability of BAO \(D_H/r_d\) observations used in GP training for reconstructing \(H(z)\). Additionally, at \(z \to 0\), \(\mathcal{O}m\) takes a \(0/0\) form, leading to numerical divergence of the function as \(z\) approaches zero. There is a noticeable narrowing of error bands at higher $z$, particularly evident around \(z = 2.33\) that requires clarification. This effect appears to be influenced by the inclusion of Ly-$\alpha$ BAO, which enhances reconstruction accuracy and precision following the BAO QSO at \(z \sim 1.49\). However, the absence of BAO data between these redshifts and the limited availability of SN-Ia beyond \(z > 1.5\) significantly compromise the reconstruction quality. As a result, broader error bars are observed in the redshift range \(1.5 < z < 2.3\), leading to reduced analytical precision in the study. These problems are even more drastic in Figure \ref{fig:Ez_vs_z_2}(b), we attribute this to the combination of Union3 SN-Ia and DESI BAO data used for GP reconstruction. The Union3 data is binned, meaning it groups data points at different redshift ranges, which reduces the detail and accuracy of the reconstruction. This reduced number of data points affects the reconstruction precision because GP works best with more data points spread out over the range being studied.

\begin{figure*}[t!]
\centering
\begin{minipage}{0.24\textwidth}
\includegraphics[width=\textwidth]{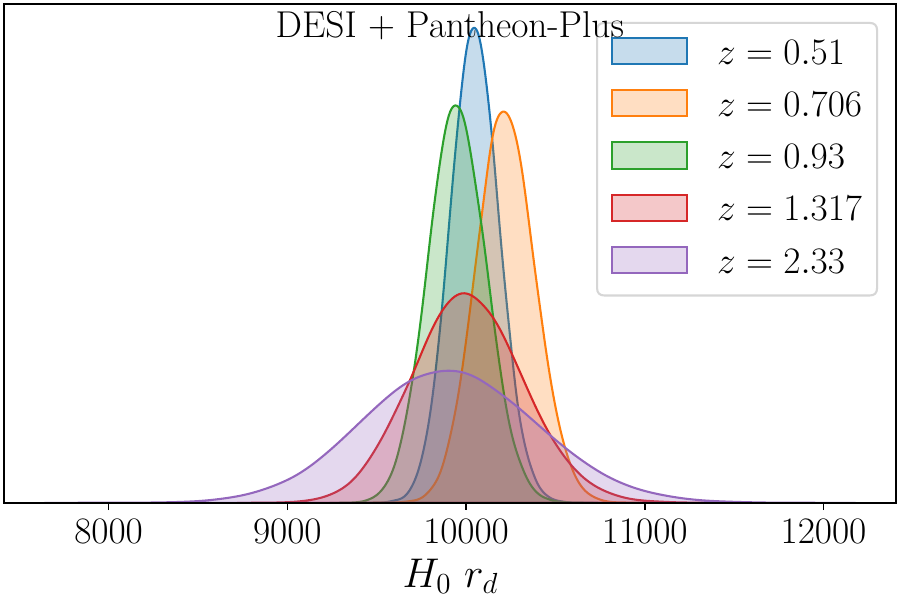}
\caption*{\small (a)} 
\end{minipage}
\begin{minipage}{0.24\textwidth}
\includegraphics[width=\textwidth]{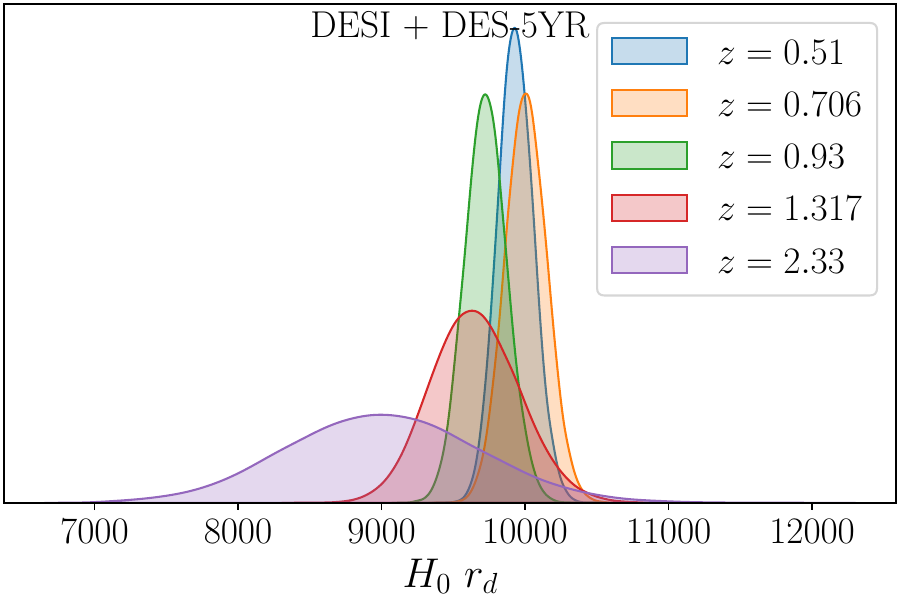}    
\caption*{\small (b)} 
\end{minipage} 
\begin{minipage}{0.24\textwidth}
\includegraphics[width=\textwidth]{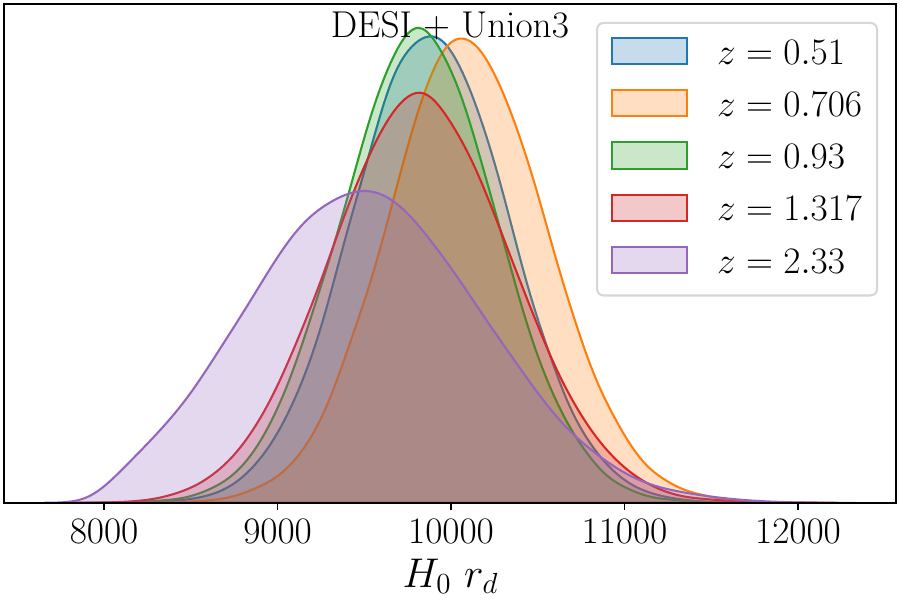}    
\caption*{\small (c)} 
\end{minipage} 
\begin{minipage}{0.24\textwidth}
\includegraphics[width=\textwidth]{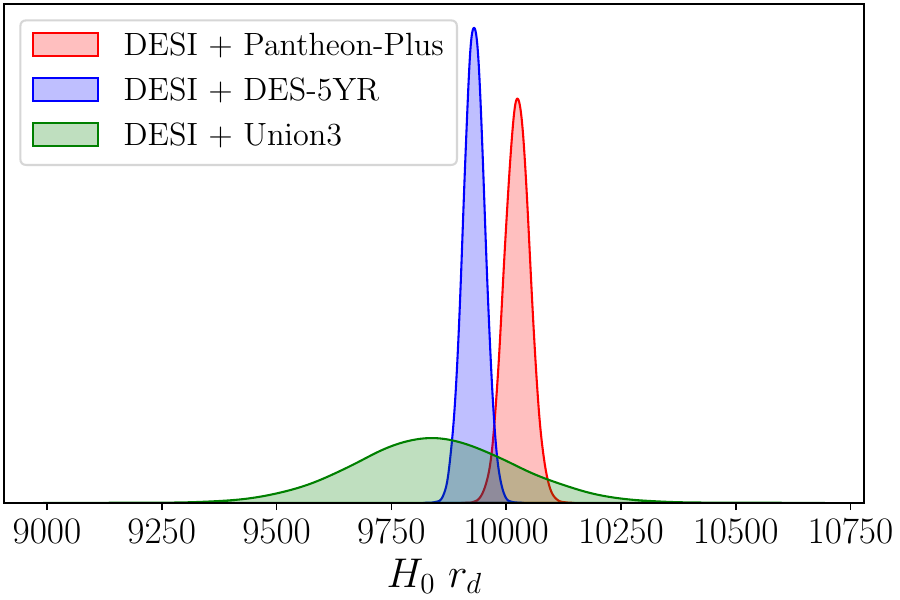}    
\caption*{\small (d)} 
\end{minipage} 
%\hfill 	
\caption{Plots for the one-dimensional marginalized $H_0 r_d$ [in km s$^{-1}$] posteriors from the (a) DESI BAO + Pantheon-Plus, (b) DESI BAO + DES-5YR and (b) DESI BAO +  Union3 data sets, at the individual DESI BAO redshift bins. (d) Plot showing a comparison between all three cases employing the full DESI BAO data in combination with SN data sets.}
\label{fig:compare_H0rd}
\end{figure*}

%\squeezetable
\begin{table}[t!]
{\renewcommand{\arraystretch}{1.25} \setlength{\tabcolsep}{15 pt} \centering
\begin{tabular}{l | c }
%        \hline 
        \hline 
        Datasets & $H_0  r_d$ [km/s] \\
        \hline \hline 
        DESI +  Pantheon-Plus & $10024.35_{-27.28}^{+26.82}$ \\
        DESI + DES-5YR & $9930.09_{-22.91}^{+22.78}$ \\
        DESI + Union3 & $9841.75_{-168.10}^{+172.11}$ \\
        \hline
%        \hline 
    \end{tabular}
    \caption{Model-independent constraints on $H_0 r_d$.}
    \label{tab:H0rd}
}
\end{table}

The next quantity we are interested in is the total or effective EoS of the Universe with an equation of state parameter $w_{\text{tot}}$, defined as $w_{\text{tot}} = \frac{p}{\rho} = -1 - \frac{2\dot{H}}{3H^2}.$ This can be rewritten as a function of the individual components, $w_{\text{tot}}= \frac{w(z) \rho_d}{\rho_m + \rho_d }$ where $w(z)$ is the dark energy EoS \cite{Mukherjee:2020ytg}. In case of $\Lambda$CDM, $w_{\text{tot}\vert \Lambda\text{CDM}} = - \left[ {\frac{\Omega_{m0}}{1-\Omega_{m0}}(1+z)^{3} + 1}\right]^{-1}$. On assuming the Planck 2018 $\Lambda$CDM best-fit $\Omega_{m0} = 0.315$ \cite{Planck:2018vyg}, we can compute $w_{\text{tot}\vert \Lambda\text{CDM}}(z=0) = -0.685$. For the same combination of datasets, we plot the evolution of $w_{\text{tot}}$ vs redshift in the bottom right panels of Fig. \ref{fig:Ez_vs_z}(a) \& Fig. \ref{fig:Ez_vs_z}(b), which involves reconstruction of the first order derivative of $E(z)$ in addition to $E(z)$. Our findings indicate that the present values of the reconstructed $w_{\text{tot}}$ are higher (i.e., less negative) than $w_{\text{tot}\vert \Lambda\text{CDM}}(z=0)$. This suggests the presence of a non-phantom EoS for DE at late times, provided that dark energy is nearly constant or very slowly varying during this late epoch of cosmic evolution, which opens up the possibility of a phantom-crossing scenario for $z \gtrsim 1$.  Our findings align with those of Yang \textit{et al}\cite{Yang:2024kdo}, who obtained an intriguing transition in $w(z)$ from the phantom to the quintessence regime as the universe expands.

Furthermore, we investigate the presence of a $\sim 2 \sigma$ discrepancy in the DESI Luminous Red Galaxy (LRG) data at $z_{\text{eff}} = 0.51$ in the redshift range $0.4 < z < 0.6$, as reported by Colg\'{a}in \textit{et al}\cite{eoin_desi}. For this exercise, we exclude the DESI LRG1 data and undertake the previous analysis using the DESI no-LRG1 + Pantheon-Plus combination and report our results in Fig. \ref{fig:compare_lrg}. Similar to the works by Dinda\cite{Dinda:2024kjf} and  Wang\cite{wang_desi}, we find that the inclusion or exclusion of DESI BAO LRG1 data hardly affects the constraints on the reconstructed functions $\mathcal{O}m$ and $w_{\text{tot}}$ obtained from the combined DESI no-LRG1 + Pantheon-Plus data set.

In the following step, we take into account two more SN-Ia compilations, namely DES-5YR \cite{DES:2024tys} and Union3 \cite{Rubin:2023ovl}, in combination with DESI BAO and compare their outcomes with those obtained while considering Pantheon-Plus. All three SN-Ia data sets are standardized to a common footing, corresponding to $H_0=70$ km Mpc$^{-1}$ s$^{-1}$. Switching from Pantheon-Plus to DES-5YR and Union3 allows for a fresh perspective on our analysis. These datasets offer different methods to check for errors and calibration systematics, potentially revealing new insights or reinforcing previous findings. Overall, this exercise enhances the reliability of our work by using diverse datasets. The plots for the reconstructed functions, $\mathcal{O}m(z)$ and $w_{\text{tot}}$ are shown in Fig. \ref{fig:Ez_vs_z_2}(a) \& Fig. \ref{fig:Ez_vs_z_2}(b) utilizing the DESI + DES-5YR and DESI + Union3 data set combinations. Results obtained utilizing the DESI + Union3 combination consistently include the Planck 2018 $\Lambda$CDM limit within the 1$\sigma$ confidence levels. However, in contrast to DESI+Pantheon-Plus and DESI+Union3,  we find that for the DESI+DES-5YR combination, the $\Lambda$CDM model assuming the Planck best-fit $\Omega_{m0} = 0.315$ is non-included within the $\mathcal{O}m$ and $w_{\text{tot}}$ reconstructions at the 3$\sigma$ confidence level in the redshift range $0.1<z<0.8$. More importantly, this behaviour remains unaffected by the choice of $M_B$ hence $H_0$ in our analysis.

Additionally, we intend to compare our results with those of the DESI Collaboration \cite{desi_paper6} to gain insights into the nature of dark energy. For this study, we refer to the marginalized posterior constraints in the $w_0-w_a$ plane for the flat $w_0,w_a$CDM model, derived from a combination of DESI BAO and SN-Ia datasets: Pantheon-Plus, DES-5YR, and Union3, as presented by Adame \textit{et al}\cite{desi_paper6} in the left panel of Fig 6. Their results conclude that each of these dataset combinations favours $w_0 > -1$ and $w_a < 0$, indicating mild discrepancies with $\Lambda$CDM at the $\geq 2\sigma$ level. There have been arguments by Cort\'{e}s \& Liddle\cite{liddle_desi} suggesting that these indications for a deviation from $\Lambda$ in favour of evolving dark energy could be strongly driven by the assumed parameter priors. Additionally, Wolf \& Ferreira\cite{Wolf:2023uno} have discussed how the $w_0-w_a$ parameterization fails to fully determine dark energy microphysics, and how conclusions and interpretations using this parameterization are highly sensitive to the redshifts probed in observational surveys. Recently, Park \textit{et al}\cite{Park:2024jns} have utilized non-DESI data to corroborate and bolster the DESI 2024 spatially-flat $w_0-w_a$CDM parameterization outcome. Despite their intrigue, these results are not yet statistically significant, as emphasized by Park \textit{et al}\cite{Park:2024jns}, who point out that the selection of $w(z)$ is merely a parameterization and not a physically consistent model of dynamical dark energy. To this end, we reiterate that the departure from $\Lambda$CDM in \cite{desi_paper6} could be an artefact imposed by the specific cosmological model being analyzed. Therefore, unlike the DESI results, it is worth noting that the selection of a specific combination of datasets (DESI+DES-5YR) influences this deviation from $\Lambda$CDM, as observed in our data-driven results.

Finally, we take a closer look at the implications of DESI BAO data through an estimation of the quantity $H_0 r_d$ in combination with the reconstructed SN-Ia data sets at individual redshift bins of the full DESI BAO ($D_M/r_d$, $D_H/r_d$ \& $D_V/r_d$) data. The one-dimensional marginalized posteriors for $H_0 r_d$ arising from the combined DESI BAO + Pantheon-Plus SN, DESI BAO + DES-5YR SN and DESI BAO + Union3 SN data is shown in Fig. \ref{fig:compare_H0rd}. Comparing the DESI + Pantheon-Plus cases with those of the DESI + DES-5YR results reveals an $\approx$ 2$\sigma$ discrepancy between the reported $H_0 r_d$ constraints, as detailed in Table \ref{tab:H0rd}. These findings are similar to those obtained by Lodha \textit{et al}\cite{DESI:2024kob} assuming well-motivated dark energy models, and Calderon \textit{et al}\cite{Calderon:2024uwn} where crossing statistics have been implemented, on the same dataset combinations. Nevertheless, our results also indicate that the individual binned $H_0 r_d$ constraints at DESI BAO effective redshifts are consistent at the 3$\sigma$ confidence level for each combination of data sets.  

\section{Concluding Remarks \label{sec:conclusion}}

In this paper, we have employed GP regression to reconstruct the evolutionary history of our Universe from current SN-Ia and BAO observations in a model-independent manner. Firstly, we have taken into consideration the latest Pantheon-Plus SN-Ia compilation together with the existing SDSS BAO and early DESI DR1 BAO measurements. The reconstructed $H(z)$ is then utilized to derive constraints on the total energy density $\rho_t(z)$, $\mathcal{O}m$ diagnostics and the total equation of state $w_{\text{tot}}$ of the Universe. We further explored the presence/absence of any anomalies in the DESI BAO LRG1 data at $z_{\text{eff}}=0.51$. We also carried out a reanalysis in light of two recently released SN-Ia compilations, namely DES-5YR and Union3, and compared their outcomes. Lastly, we examined the $H_0 r_d$ constraints arising from the DESI BAO and SN-Ia data sets, at the BAO effective $z$-bins, and studied the possibility of running scenarios in the $H_0 r_d$ parameter at different $z$-values. This exercise has been complemented by analysing the mutual consistency between the three compilations of SN-Ia and DESI BAO. Our findings have been summarized as follows:
\begin{itemize}
    \item The reconstructed functions $E(z)$, $\mathcal{O}m(z)$ and $w_{\text{tot}}(z)$ from SDSS vs DESI BAO in combination with Pantheon-Plus SN-Ia include $\Lambda$CDM within the 2$\sigma$ CL. This indirectly implies the mutual consistency between SDSS BAO and DESI BAO measurements. The inclusion or exclusion of the DESI BAO LRG1 data does not affect our final results.

    \item Reconstructions with DESI+DES-5YR, which predict the exclusion of $\Lambda$CDM at 3$\sigma$ in the range $0.1<z<0.8$, reveals a tension with the results obtained with DESI+Pantheon-Plus. However, results obtained using the DESI+Union3 combination always include the Planck 2018 $\Lambda$CDM limit within the 1$\sigma$ CL, suggesting a need for further reassessment of the DES-5YR data. 

    \item The binned $H_0 r_d$ constraints at DESI BAO effective redshifts are consistent at the 3$\sigma$ confidence level for all three data combinations- DESI+Pantheon-Plus, DESI+DES-5YR and DESI+Union3, separately. However, the $H_0 r_d$ constraints from the DESI+Pantheon-Plus are in $\approx$ 2$\sigma$ tension with the DESI+DES-5YR results.    

    \item The reconstructed $\mathcal{O}m(z)$ for all the cases exhibits a characteristic negative slope for $z<1$, and the derived constraints on $w_{\text{tot}}$ are greater than $w_{\text{tot}\vert \Lambda\text{CDM}}$ at the present epoch, suggests that dark energy is quintessence-like with a nearly constant or very slowly varying EoS during this late epoch of cosmic evolution. The possibility of a phantom-crossing scenario for $z \gtrsim 1$ cannot be excluded with the DESI BAO and SN-Ia combinations.

    \item The conclusions regarding the 2$\sigma$ departure from $\Lambda$CDM with DESI BAO could be an artefact imposed by the $w_0 w_a$-model being analyzed by the DESI Collaboration \cite{desi_paper6}. We find that the selection of a specific combination of datasets (DESI+DES-5YR) influences this deviation from $\Lambda$CDM.

    \item Addressing the $H_0$ tension calls for either modifications to the late-time or early Universe physics, or search for hidden systematics in the data sets. Our results confirm that SN-Ia and BAO observations do not rule out any of the three possibilities.

\end{itemize}

Finally, given the status of the current results, it will be interesting to explore the applicability of alternative machine learning techniques besides GP, such as genetic algorithms \cite{Nesseris:2012tt, Arjona:2021mzf, Bernardo:2021mfs, Gangopadhyay:2023nli} and neural networks \cite{Shah:2024slr, Dialektopoulos:2023dhb, Gomez-Vargas:2022bsm, Gomez-Vargas:2021zyl} for reconstructing the evolutionary history in a cosmological model-independent way to draw inferences on the validity of the vanilla $\Lambda$CDM model.

\begin{acknowledgements}
\noindent We thank the anonymous reviewer for their valuable suggestions towards the improvement of the manuscript. PM thanks Eoin \'{O} Colg\'{a}in, Supratik Pal and Rahul Shah for interesting discussions. PM acknowledges funding from the Anusandhan National Research Foundation (ANRF), Govt of India under the National Post-Doctoral Fellowship (File no. PDF/2023/001986). AAS acknowledges the funding from ANRF, Govt of India under the research grant no. CRG/2020/004347. We acknowledge the use of HPC facility, Pegasus, at IUCAA, Pune, India.
\end{acknowledgements}

%%%%%%%%%%%%%%%%%%%%%%%%%%%%%%%%%%%%%%%

\appendix

\onecolumngrid
\section{Gaussian Process Regression \label{sec:gp}}

For a given set of training inputs ${X_1} = \left\lbrace x_{1,i}\right\rbrace $ and ${X_2} = \left\lbrace x_{2,i}\right\rbrace $, our dataset includes observations of $f(X_1)$ versus $X_1$ and its derivative $f^\prime(X_2)$ versus $X_2$, along with their respective covariance matrices ($\text{cov}[f, f]$, $\text{cov}[f^\prime, f^\prime]$, and $\text{cov}[f, f^\prime]$). We employ GPR to predict $f(X^{\star})$ and $f^\prime(X^{\star})$ on a test set $X^{\star} = {x_{i}^{\star}}$. Assuming the observed data, including noise, and the predicted functions follow a multivariate normal distribution, the reconstructed functions $f(x)$ and its derivative $f^\prime(x)$ at any point $x$ are characterized by some mean $\mu(x) = \mathbb{E}[f(x)]$, $\mu^\prime(x) = \mathbb{E}[f^\prime(x)]$ and covariance function (or kernel) $\text{cov}[f(x), f(x')] = \kappa(x, x')$, which describes the correlation between function values at points $x$ and $x'$. This study utilizes several kernels found in the literature \cite{10.7551/mitpress/3206.001.0001, Seikel:2012uu}, namely
\begin{enumerate}
\item Squared Exponential (RBF): $\kappa(x, \tilde{x}) = \sigma_f^2 \exp{\left[ - \frac{(x - \tilde{x})^2}{2l^2}\right]}$,
\item Rational Quadratic (RQD): $\kappa(x, \tilde{x}) = \sigma_f^2 \left[ 1 + \frac{(x - \tilde{x})^2}{2\alpha l^2}\right]^{-\alpha}$,
\item Matérn with $\nu = 7/2$ and $9/2$ (M72 \& M92): $\kappa(x, \tilde{x}) = \sigma_f^2 \frac{\left[ \frac{\sqrt{2 \nu}}{l} (x - \tilde{x})\right]^\nu}{\Gamma(\nu 2^{\nu -1})} K_{\nu} \left( \frac{\sqrt{2\nu}}{l}(x-\tilde{x})\right)$,
\end{enumerate}
where, $\nu$ denotes the order, $K_\nu(\cdot)$ and $\Gamma(\cdot)$ refer to the modified Bessel and Gamma functions, respectively. The kernel hyperparameters, $\sigma_f$, $l$, and $\alpha$, regulate the amplitude of fluctuations, the length scale of correlations, and the balance between large-scale and small-scale variations across different distances between points. 

Suppose, $D= \left[f(X_1) \,~ f^\prime(X_2) \right]^\text{T}$ denotes joint training dataset, the prior covariance matrix between the training points $X_1$ \& $X_2$ is  given by
\begin{equation}
    K = \begin{bmatrix}
\kappa(X_1,X_1) & \kappa^\prime(X_1,X_2) \\
\kappa^\prime(X_1,X_2)^\text{T} & \kappa^{\prime\prime}(X_2,X_2)
\end{bmatrix}  \, .
\end{equation} 
Here $\kappa^\prime(x_i, x_j) = \frac{\partial \kappa(x_i, x_j)}{\partial x_j}$ and $\kappa^\prime(x_i, x_j) = \frac{\partial^2 \kappa(x_i, x_j)}{\partial x_i \partial x_j}$ respectively. The full noise matrix of the data is 
\begin{equation}
    \mathcal{C} = \begin{bmatrix}
    \text{cov}[f(X_1),f(X_1)] & \text{cov}[f(X_1),f^\prime(X_2)] \\
\text{cov}[f(X_1),f^\prime(X_2)]^\text{T} & \text{cov}[f^\prime(X_2),f^\prime(X_2)]
\end{bmatrix} \, .
\end{equation}
The predicted values of the underlying function $f$ and its derivatives $f^\prime$ are given by,
\begin{equation}
     \begin{bmatrix}
         f(X^\star) \\
        f^\prime(X^\star)  
     \end{bmatrix} =      \begin{bmatrix}
         \mu(X^\star) \\
        \mu^\prime(X^\star)  
     \end{bmatrix} + A \, \left[ K + \mathcal{C} \right]^{-1} \, D \, , 
\end{equation}
where \begin{equation}
A = \begin{bmatrix}
        \kappa(X^\star , X_1) \,~  \kappa^\prime(X^\star, X_2) \\
        \kappa^\prime(X^\star , X_1) \,~  \kappa^{\prime\prime}(X^\star, X_2) \\
    \end{bmatrix} \, . 
\end{equation}
Again, the resulting covariance between the predictions are
\begin{align}
\text{cov}[f(X^\star), f(X^\star)] &=  \kappa(X^\star, X^\star) - \left[\kappa(X^\star, X_1) \,~ \kappa^\prime(X^\star, X_2) \right] K^{-1} \, \left[\kappa(X^\star, X_1) \,~ \kappa^\prime(X^\star, X_2) \right]^\text{T} \, , \\
\text{cov}[f^\prime(X^\star), f^\prime(X^\star)] &=  \kappa^{\prime\prime}(X^\star, X^\star) - \left[\kappa^\prime(X^\star, X_1) \,~ \kappa^{\prime\prime}(X^\star, X_2) \right] K^{-1} \, \left[\kappa^\prime(X^\star, X_1) \,~ \kappa^{\prime\prime}(X^\star, X_2) \right]^\text{T} \, , \\ 
\text{cov}[f(X^\star), f^\prime(X^\star)] &=  \kappa^\prime(X^\star, X^\star) - \left[\kappa(X^\star, X_1) \,~ \kappa^\prime(X^\star, X_2) \right] K^{-1} \, \left[\kappa^\prime(X^\star, X_1) \,~ \kappa^{\prime\prime}(X^\star, X_2) \right]^\text{T} \, . 
\end{align}
To ensure a cosmological model-independent analysis, we assume zero mean [$\mu(X^\star)=\mu^\prime(X^\star)=0$] as prior information for predictions. Hence, the choice of kernel determines almost all the generalization properties of our GPR. The hyperparameters are obtained by maximizing the log-marginal likelihood
\begin{equation} \label{eq:loglike}
\ln \mathcal{L}\left( \sigma_f , l , M_B, \cdots \right) = -\frac{1}{2} D^\text{T} \left[ K + \mathcal{C}\right]^{-1} D - \frac{1}{2} \ln \vert K + \mathcal{C} \vert - \frac{n_1 + n_2}{2} \ln (2\pi) \, ,
\end{equation}
where $n_1$ and $n_2$ are the size of the training points $X_1$ and $X_2$ respectively. For our Bayesian analysis, we marginalize the loglikelihood (as shown in Fig. \ref{fig:hyperplot}) assuming flat priors on the kernel hyperparameters. The training points $X_1$ and $X_2$ are respectively the redshifts $z$ of SN-Ia and BAO observations. The data vector $D$ comprises of the SN-Ia and BAO $D_M(z)$ and $D_H(z)$ measurements. The kernel hyperparameters ($\sigma_f$, $l$) together with the cosmological parameters viz. the supernovae absolute $M_B$ magnitude or additional parameters governing the GP mean function, are marginalised over for the maximum log-likelihood analysis employing MCMC methods.

\subsection{Kernel Selection}

To determine the most suitable kernel we select the GP model that gives the minimum $\chi^2$ between the observations vs reconstructed values at the training points. We compute the reduced $\chi^2_{\text{min}}$, denoted as \(\tilde{\chi}^2_{\text{min}}\), where \(\tilde{\chi}^2_{\text{min}} = \chi^2_{\text{min}} / \text{d.o.f}\). Here, `d.o.f' represents the number of degrees of freedom, calculated as the number of data points minus the number of hyperparameters \cite{Favale:2023lnp}. Table \ref{tab:kernel} gives the best-fit $\chi^2_{\text{min}}$ for the four kernels, which shows that the Mat\'{e}rn $\nu = 7/2$ (M72) kernel performs slightly better than the others. However, we got closely similar results irrespective of the kernel choice \cite{Banerjee:2023evd, Banerjee:2023rvg, Favale:2023lnp}.

\subsection{Effect of Mean Function}

\begin{figure*}[t!]
\centering
\begin{minipage}{0.26\textwidth}
\includegraphics[width=\textwidth]{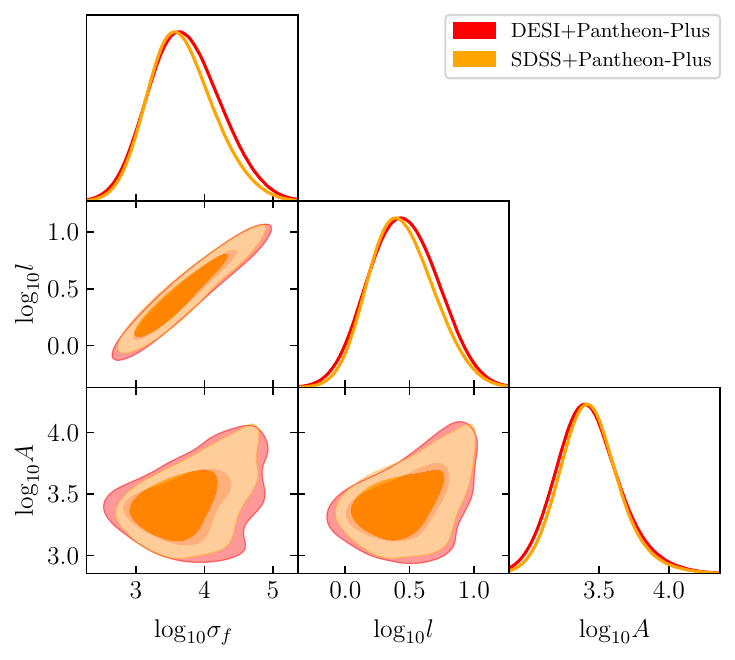}
\caption*{\small (a)} 
\end{minipage} 
\begin{minipage}{0.36\textwidth}
\includegraphics[width=\textwidth]{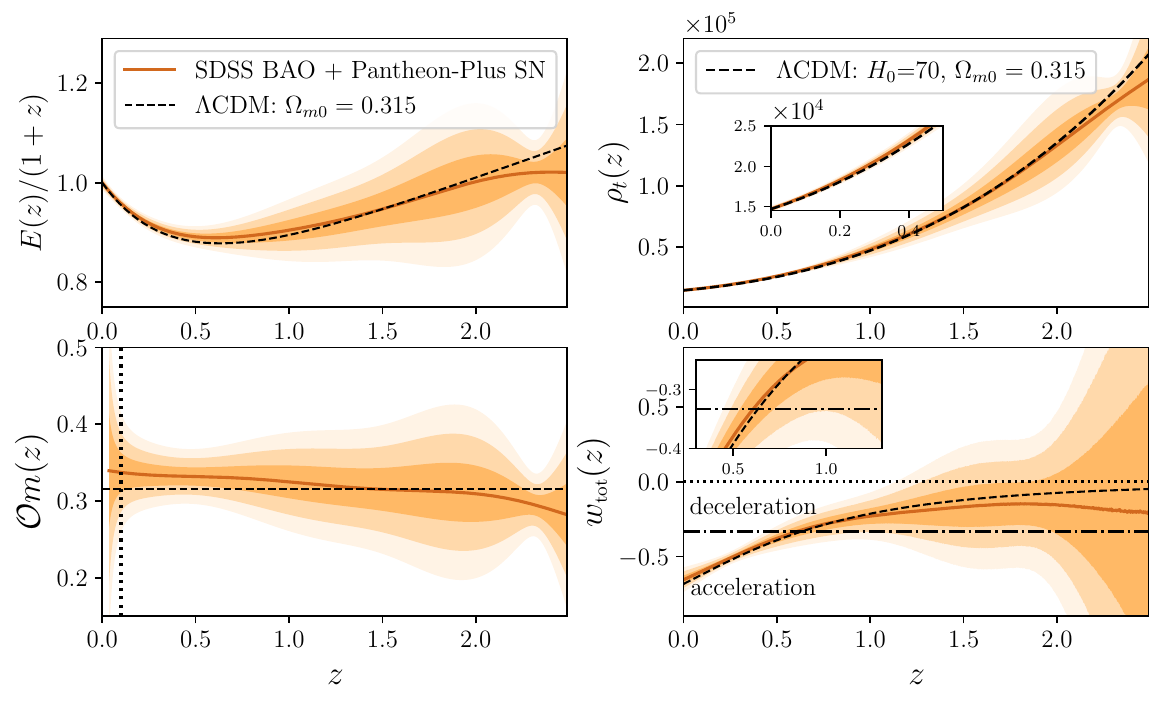}
\caption*{\small (b)} 
\end{minipage}
\begin{minipage}{0.36\textwidth}
\includegraphics[width=\textwidth]{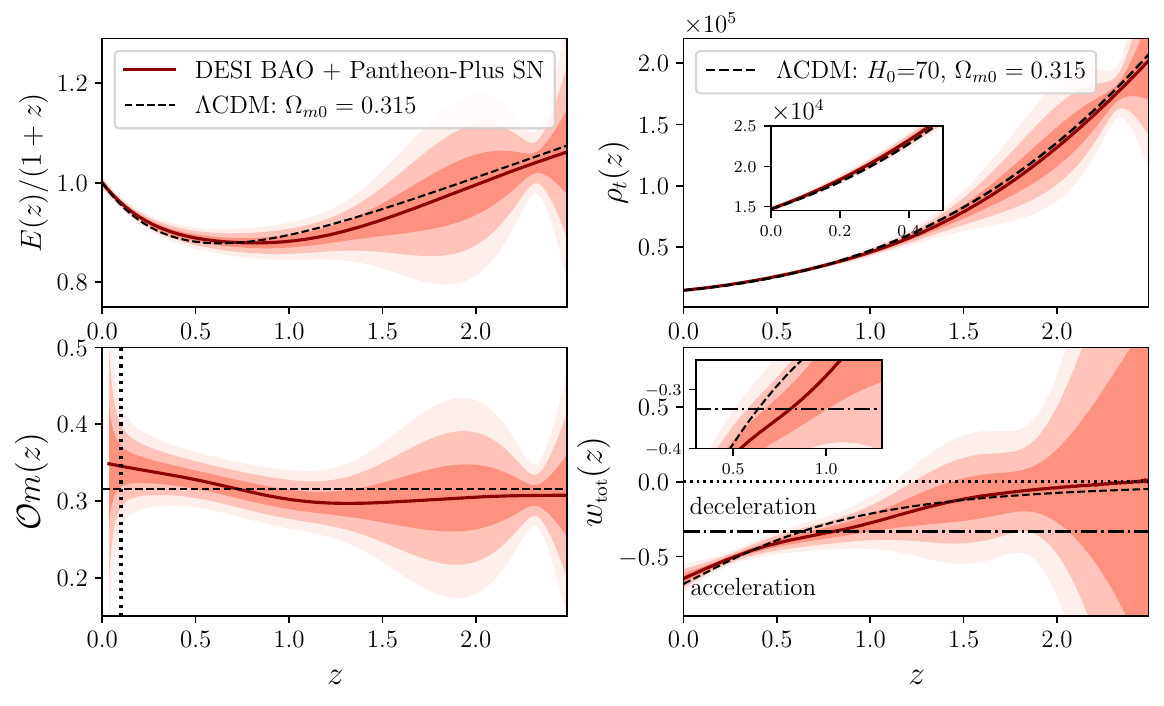}    
\caption*{\small (c)} 
\end{minipage} 
%\hfill 	
\caption{(a) Contour plots for GP hyperparameter space and parameters governing the GP mean function. Plots for mean values of the reconstructed functions along with the 1$\sigma$, 2$\sigma$ and 3$\sigma$ uncertainties for the (a) SDSS+Pantheon-Plus vs (b) DESI+Pantheon-Plus data, employing a linear-in-redshift mean function for $D_M(z)$. }
\label{fig:mean_plot}
\end{figure*}

There have been ongoing discussions about the implications of using a zero mean function, particularly regarding small length scales in the kernel and for dynamical quantities like distances in cosmology. The choice of the mean function is crucial and should carefully be selected, followed by a full marginalisation over the parameters associated with this mean function together with the kernel hyperparameters. This reduces the dependence on the input mean function - although not exhaustive - as there are limitations and assumptions inherent in this approach too, and also requires proper selection for this mean function, which in turn leads to additional modelling \cite{Hwang:2022hla}.

In light of these concerns, we now investigate the effect of using a non-zero linear-in-redshift mean function, \(\mu(z) = A \, z\), as an additional prior in modeling the cosmic comoving distance \(D_M(z)\) for our GP reconstruction. In this framework, the log-marginal likelihood is expressed as follows:

\[
\ln \mathcal{L}\left( \sigma_f, l, M_B, A, \cdots \right) = -\frac{1}{2} (D - M )^\text{T} \left[ K + \mathcal{C}\right]^{-1} (D - M) - \frac{1}{2} \ln \vert K + \mathcal{C} \vert - \frac{n_1 + n_2}{2} \ln (2\pi) \, ,
\]

where \(M = \left[\mu(X_1) \,~ \mu^\prime(X_2) \right]^\text{T}\) represents the joint mean function. We adopt a non-informative uniform prior for \(A\), specifically \(\log_{10} A \in [-5, 6]\), alongside the other priors detailed in Table \ref{tab:priors}.

The resulting two-dimensional parameter space for the GP hyperparameters and the mean function parameter \(A\) is illustrated in Fig. \ref{fig:mean_plot}(a), using the SDSS+Pantheon-Plus and DESI+Pantheon-Plus datasets. The final reconstruction profiles for \(E(z)\), \(\rho_t(z)\), \(\mathcal{O}m(z)\), and \(w_{\rm tot}(z)\) (similar to those shown in Fig. \ref{fig:Ez_vs_z}) are presented in Figs. \ref{fig:mean_plot}(b) and \ref{fig:mean_plot}(c) as functions of redshift \(z\). The reduced \(\chi^2_{\rm min}\) values for both combinations are \(0.875\) for SDSS+Pantheon-Plus and \(0.879\) for DESI+Pantheon-Plus, which are comparable to the values obtained with the zero-mean function (see Table \ref{tab:kernel}) when employing the Mat\'{e}rn 7/2 kernel. Our findings demonstrate that both the zero mean and the linear-in-\(z\) mean function choices yield similar predictive performances and generalization properties for the reconstructed functions, such as allowing $\Lambda$CDM within 2$\sigma$, presence of quintessence-like nature of dark energy at present epoch as well as the possibility for phantom-crossing scenario at $z \gtrsim 1$.

\section{Observational Datasets  \label{sec:data}}

\subsection{Baryon Acoustic Oscillations}

Our analysis focuses primarily on the Dark Energy Spectroscopic Instrument (DESI) 2024 Baryon Acoustic Oscillation (BAO) observations \cite{desi_paper6}. These observations encompass five key samples: the Luminous Red Galaxy samples (LRG) at effective redshifts \( z_{\text{eff}} = 0.51 \) and \( z_{\text{eff}} = 0.71 \) \cite{DESI:2022gle}, combinations of LRG and the Emission Line Galaxy sample (ELG) at \( z_{\text{eff}} = 0.93 \) \cite{desi_paper6}, the standalone ELG sample at \( z_{\text{eff}} = 1.32 \) \cite{Raichoor:2022jab}, and the Lyman-$\alpha$ forest sample (Ly-$\alpha$) at \( z_{\text{eff}} = 2.33 \) respectively \cite{desi_paper6}.

In addition to the DESI DR1 BAO, we include the completed Sloan Digital Sky Survey (SDSS) IV 2020 BAO observations \cite{eBOSS:2020yzd}, which comprise five pairs of measurements of \( D_M/r_d \) and \( D_H/r_d \) at effective redshift bins. The first and second pairs correspond to the Baryon Oscillation Spectroscopic Survey (BOSS) DR12 LRG sample at \( z_{\text{eff}} = 0.38 \) and 0.51, respectively \cite{BOSS:2016wmc}. The third pair corresponds to the extended Baryon Oscillation Spectroscopic Survey (eBOSS) DR16 LRG sample at \( z_{\text{eff}} = 0.698 \) \cite{eBOSS:2020yzd}. The fourth pair corresponds to the eBOSS DR16 QSO sample at \( z_{\text{eff}} = 1.48 \) \cite{eBOSS:2020gbb}. Lastly, the fifth pair corresponds to the eBOSS DR16 Ly-$\alpha$ QSO sample at \( z_{\text{eff}} = 2.334 \) \cite{eBOSS:2020tmo}.

\subsection{Type Ia Supernovae}

We employ three distinct SN-Ia datasets in our analysis. The Pantheon-Plus compilation \cite{Scolnic:2021amr} includes 1550 spectroscopically-confirmed SN-Ia spanning the redshift range $0.001 < z < 2.26$. For this dataset, we incorporate comprehensive statistical and systematic covariance and apply a bound of $z > 0.01$ to minimize the influence of peculiar velocities on the Hubble diagram \cite{Peterson:2021hel}. The recently released Union3 compilation \cite{Rubin:2023ovl} comprises 2087 SN-Ia, with 1363 SN-Ia overlapping with Pantheon-Plus, which employs a distinct approach to handling systematic errors and uncertainties through Bayesian Hierarchical Modelling. Additionally, the Dark Energy Survey Year 5 data release (DES5YR) presents a newly selected sample of 1635 photometrically-classified SN-Ia with redshifts ranging from $0.1 < z < 1.3$, complemented by 194 low-redshift SN-Ia (overlapping with Pantheon-Plus) in the range $0.025 < z < 0.1$ \cite{DES:2024tys, DES:2024upw, the_des_sn_working_group_2024_12720778}. We avoid combining the results of these SN-Ia datasets since they are not mutually exclusive\footnote{Pantheon-Plus and Union3 share approximately 1360 supernovae but differ in their methodology for analyzing and marginalizing astrophysical and systematic parameters. DES5YR contributes a new high-$z$ dataset of around 1500 photometrically classified SN-Ia, while still incorporating approximately 194 historical SN-Ia at $z < 0.1$ that overlap with the other two datasets.}.

\twocolumngrid

\bibliographystyle{apsamp4-2}
\bibliography{references.bib}

\end{document}